\documentclass[%
 aip,
 amsmath,amssymb,
 reprint,%
]{revtex4-1}

\usepackage{graphicx}
\usepackage{dcolumn}
\usepackage{bm}

\usepackage[table,x11names]{xcolor}
\usepackage{graphicx}		
\usepackage{dcolumn}		
\usepackage{graphicx}		
\usepackage{dcolumn}		
\usepackage{makecell}
\usepackage{textcomp}
\usepackage{xfrac}
\usepackage{xcolor} 

\usepackage[hidelinks]{hyperref}
\usepackage{url}
\usepackage{comment}
\usepackage{floatrow}
\usepackage{commath}
\usepackage{amsmath}
\usepackage[]{lineno}
\modulolinenumbers[1]

\begin{document}

\preprint{AIP/123-QED}

\title{Multiple Monoenergetic Gamma Radiography (MMGR) with a compact superconducting cyclotron}
\author{Hin Y. Lee}
\author{Brian S. Henderson}
\altaffiliation[Currently with ]{the MITRE Corporation, Bedford, Massachusetts 01730, USA.  The author’s affiliation with the MITRE Corporation is provided for identification purposes only and is not intended to convey or imply MITRE’s concurrence with, or support for, the positions, opinions, or viewpoints expressed in this work.}
\author{Roberts G. Nelson}
\author{Areg Danagoulian}
\email[E-mail: ] {aregjan@mit.edu}
\affiliation{Department of Nuclear Science and Engineering, Massachusetts Institute of Technology, Cambridge, Massachusetts 02139, USA}
\date{\today}

\begin{abstract}
Smuggling of special nuclear materials (SNM) and nuclear devices through borders and ports of entry constitutes a major risk to global security.  Technologies are needed to reliably screen the flow of commerce for the presence of high-$Z$ materials such as uranium and plutonium.  Here we present an experimental proof-of-concept of a technique which uses inelastic ($p,p'$) nuclear reactions to generate monoenergetic photons, which provide means to measure the areal density and the effective-$Z$ ($Z_{\text{eff}}$) of an object with an accuracy which surpasses that achieved by current methods. We use an ION-12$^{ \text{SC}}$ superconducting 12~MeV proton cyclotron to produce 4.4, 6.1, 6.9, and 7.1~MeV photons from a variety of nuclear reactions.    Using these photons in a transmission mode we show that we are able to accurately reconstruct the areal densities and $Z_{\text{eff}}$ of a test object.  This methodology could enable mobile applications to screen commercial cargoes with high material specificity, providing a means of distinguishing common cargo materials from high-Z materials that include uranium and plutonium.

\end{abstract}
\maketitle{}


\section{Introduction}
\label{sec:MMGR}

Within the field of nuclear security cargo security focuses on the screening of the flow of commerce for the presence of special nuclear materials (SNM) (such as plutonium and uranium), fully assembled weapons, as well as other non-nuclear contraband.  The detection of SNM in particular is important in the context of nuclear terrorism. It is estimated that the economic impact alone in case of the detonation of a crude nuclear weapon in an urban setting can cause damage in excess of \$1~trillion~\cite{abtes}.  As many as 50,000 maritime intermodal containers  enter the United States daily through a variety of maritime ports and other pathways~\cite{kouzes}.  Given the density of their content \cite{henderson2019analysis}, their anonymity, and the speed at which these need to be screened, screening cargo containers for nuclear materials presents a major challenge to national and global security.  In this paper we present a new radiographic concept that uses transmission of photons of precise and well known energies produced by a superconducting cyclotron  to achieve accurate determination of an object's areal density and material type. 

Modern passive detection systems in ports of entry have the capability to detect the radiation signals from spontaneous nuclear decay from some types of SNM. However, these signals have a low signal-to-background ratio when the source is shielded, rendering passive detection mostly ineffective against well-informed smugglers.  In particular, $^{235}$U has a low spontaneous fission rate and has only intense gamma lines at 186~keV and lower energies, which are easy to shield. This makes highly enriched uranium difficult to detect via passive methods. Therefore, scanning of cargoes that potentially contain shielded SNM requires active interrogation and transmission radiography techniques that utilize external radiation sources.  Furthermore, while screening for radiation may reveal some SNM, it does not address the auxiliary --- and equally important for the transportation industry and customs agencies --- goal of detecting other types of contraband, such as smuggling of economic goods.  Active interrogation involves directing nuclear radiation such as photons and neutrons to the object and subsequently measuring the secondary radiation, such as delayed and prompt neutrons from photofission to gain information about the content of the object~\cite{pnpf-short, mayer, nattress2019high}. Transmission radiography involves measuring the attenuated flux of the interrogating beam's transmitted particles, comparing it with the incident flux, and using that to infer the areal density and possibly the material type subtending a particular pixel. An additional strength of radiography, when compared to other active methods, is its ability to achieve imaging and thus allow detection of conventional contraband - which is the primary goal for many customs agencies.  Most traditional X-ray radiography methods make use of bremsstrahlung beams:  the polychromatic nature of these beams however are their main limitations, as the lack of energy-specificity reduces their sensitivity to the $Z$ of the cargo. Monoenergetic beams are more advantageous as the gammas have well defined energies, allowing for much better sensitivity to $Z$ and thus improved reconstructions, as can be seen for example in the work of~\citeauthor{brian}\cite{brian}.
While radiographic techniques are primarily based on photons, techniques involving other particles do exist such as using neutron beams~\cite{Rahon,rahon2020hydrogenous}. For a high-level discussion of other active interrogation methods and bremsstrahlung radiography see \citeauthor{runkle_rattling}~\cite{runkle_rattling} and Section 2.2.6 in~\citeauthor{JimmyPhD}\cite{JimmyPhD}. 

Prior research on monoenergetic gamma radiography was conducted at the MIT Bates Research and Engineering Center and has been described in detail by \citeauthor{brian}\cite{brian} and \citeauthor{rose2016uncovering}\cite{rose2016uncovering}.  Similar concepts have been described by~\citeauthor{goldberg2008method}\cite{goldberg2008method,goldberg2010dual}. All these approaches however had a number of significant disadvantages and limitations: a large ($\sim$3 meters long) radiofrequency quadrupole (RFQ) accelerator requiring additional bulky supporting hardware; a pulsed beam, which in some cases caused significant detector pileup; and the use of $^{11}$B($d,n\gamma$)C$^{12}$ reaction which produced more neutrons than photons.  The large size limits RFQ's fieldability, while the neutrons contribute to radiation dose. In this work we instead use a more fieldable accelerator platform and a different type of a nuclear reaction.  The experiment uses a compact, relatively light and low-power superconducting cyclotron of a size of $2.5 \times2 \times 2$ m$^3$.  For an in-depth description see \citeauthor{IONETIX_paper} \cite{IONETIX_paper} and \citeauthor{IONETIX_patent}\cite{IONETIX_patent}.  Using photons from the neutron-less $^{\text{12}}$C ($p,p'\gamma$)$^{\text{12}}$C and $^{\text{16}}$O($p,p'\gamma$)$^{\text{16}}$O reactions we show that it is possible to accurately reconstruct the areal density $x$ and effective atomic number $Z_{\text{eff}}$ of a variety of homogeneous and heterogeneous objects with  $Z_{\text{eff}}$ values in the range of 13--92.  While the cyclotron has its own limitations, in particular in the form of an internal target that is difficult to cool and shield, it is a significant step forward in terms of applicability.

 The attenuation of photons in materials is approximated by the following equation: 
\begin{gather}  
\label{eq:total_attenuation}
A=\frac{I}{I_0}=e^{-\mu x}~~~\text{where}~~~\mu = \mu_{\text{PE}} + \mu_{\text{CS}} + \mu_{\text{PP}}
\end{gather}
 where $I$ is the transmitted photon intensity, $I_0$ is the source intensity, $x$ is the areal density, and $\mu$ is the total mass attenuation coefficient from the photoelectric effect~(PE), Compton scattering~(CS), and $e^+e^-$ pair production~(PP). The parameters $\mu_{\text{PE}}$, $\mu_{\text{CS}}$, $\mu_{\text{PP}}$ are the mass attenuation coefficients for PE, CS, and PP, respectively. As indicated by Eq.~\ref{eq:total_attenuation} radiography at a single energy depends on both $x$ and $\mu$, the latter being a function of both energy and $Z$, resulting in an underdetermined equation of multiple unknowns.  Measurements at multiple energies may allow for simultaneous determination of $x$ and $\mu$, the latter allowing $Z$ to be inferred.  As the number and energy separation of the lines increases, the accuracy of the reconstruction improves as well.  One method to determine $Z_{\text{eff}}$ and $x$ of the cargo material is to exploit the $Z$ and energy dependence of $\mu$ and to take transmission measurements at two or more energies~\cite{buck,oday2015,brian}. Above $4$~MeV the major photon interaction mechanisms are CS and PP, and their contributions to the mass attenuation coefficient can be described as follows~\cite{brian,leo}:
\begin{gather*}
\mu_{\text{CS}} =  N_A \sigma_{\text{CS}}(E,Z) Z/ A \\
\mu_{\text{PP}} = N_A \sigma_{\text{PP}}(E,Z) / A \\
\sigma_{\text{CS}} \propto 1/E ~\text{for}~E\gg m_e \\
\label{eq:pp}
\sigma_{\text{PP}} \propto Z^2 f(E) 
\end{gather*}
 where $\mu_{\text{CS}}$ and $\mu_{\text{PP}}$ are the mass attenuation coefficient of CS and PP with the corresponding interaction cross sections $\sigma_{\text{CS}}$ and $\sigma_{\text{PP}}$, $N_A$ is  Avogadro's number, $A$ is the atomic number of the material, $m_e$ is the rest mass of electron, and $E$ is the energy of the photon. As discussed in \citeauthor{brian}, the $f(E)$ term in the pair production cross section estimation is a function of energy with negligible dependence on $Z$ except for the Bohr correction. Since the $Z/A$ ratio for most isotopes is in the range  0.4--0.5, the attenuation from CS does not significantly depend on $Z$.  It primarily depends on the areal density $x$ and energy of the photon. Similarly, the mass attenuation coefficient of pair production is linearly dependent on the $Z$ value of the material.  
 CS and PP dominate at different energy ranges. For high-$Z$ materials  CS dominates from 1 to 3.5~MeV, hence in that range the attenuation is primarily dependent on areal density. PP is the dominant interaction above 5~MeV.   See \citeauthor{knoll}, Figure~2.20 for a more detailed description.   A measurement at lower energies can allow for a determination of areal density, with a subsequent measurement at higher energies allowing to infer $Z$ -- as shown by \citeauthor{brian}, Eq.~2, the measurement of attenuation at two energies with clearly dominating processes could be used to infer $Z$: $\ln(A(E_\text{PP}))/\ln(A(E_\text{CS})) \propto Z$. 
 The above is a schematic description of the approach, with the only goal of describing the overall physics behind it.  Since the lowest energy line used in this study is 4.4~MeV, where both CS and PP are important, a sophisticated reconstruction analysis is required, as described in Section~\ref{sec:res}.

\section{Experimental Methods}
\subsection{Compact superconducting cyclotron for MMGR}
 
 The accelerator used in this study is a 12~MeV \mbox{ION-12}$^{\text{SC}}$ isochronous proton cyclotron made by Ionetix~\cite{IONETIX_paper}. It has an internal target and is designed to produce the $^{\text{13}}$N radioisotope for positron emission tomography.  The cyclotron operates in continuous wave mode, which significantly reduces pulse pileup in the detectors. To create the needed photons, the internal target uses a graphite collimator and a static water pocket with a 50~{\textmu}m-thick aluminum window. The impinging proton beam creates photons through ($p,p^{\prime}\gamma$) reactions. Since the energy threshold of the $^{12}$C($p,n$)$^{12}$N and $^{16}$O($p,pn$)$^{15}$O reactions are 19.64 and 16.65~MeV respectively,  the only sources of neutrons are the ($p,n$) and ($p,pn$) reactions on aluminum~\cite{qcalc}. The simplicity of the target design was behind the choice of aluminum as a window material.  Using the 50~{\textmu}m-thick aluminum target window the observed neutron dose near the detectors was 300~{\textmu}Rem/hr. The neutron dose can be further reduced by using other materials: a separate measurement with a 125~{\textmu}m-thick Kapton window showed a reduction of the dose to just 26~{\textmu}Rem/hr. The aluminium window was however used for the experiment due to its durability and resistance to radiation damage.
 
\begin{figure}[thb]
    \centering
    \includegraphics[width=\columnwidth]{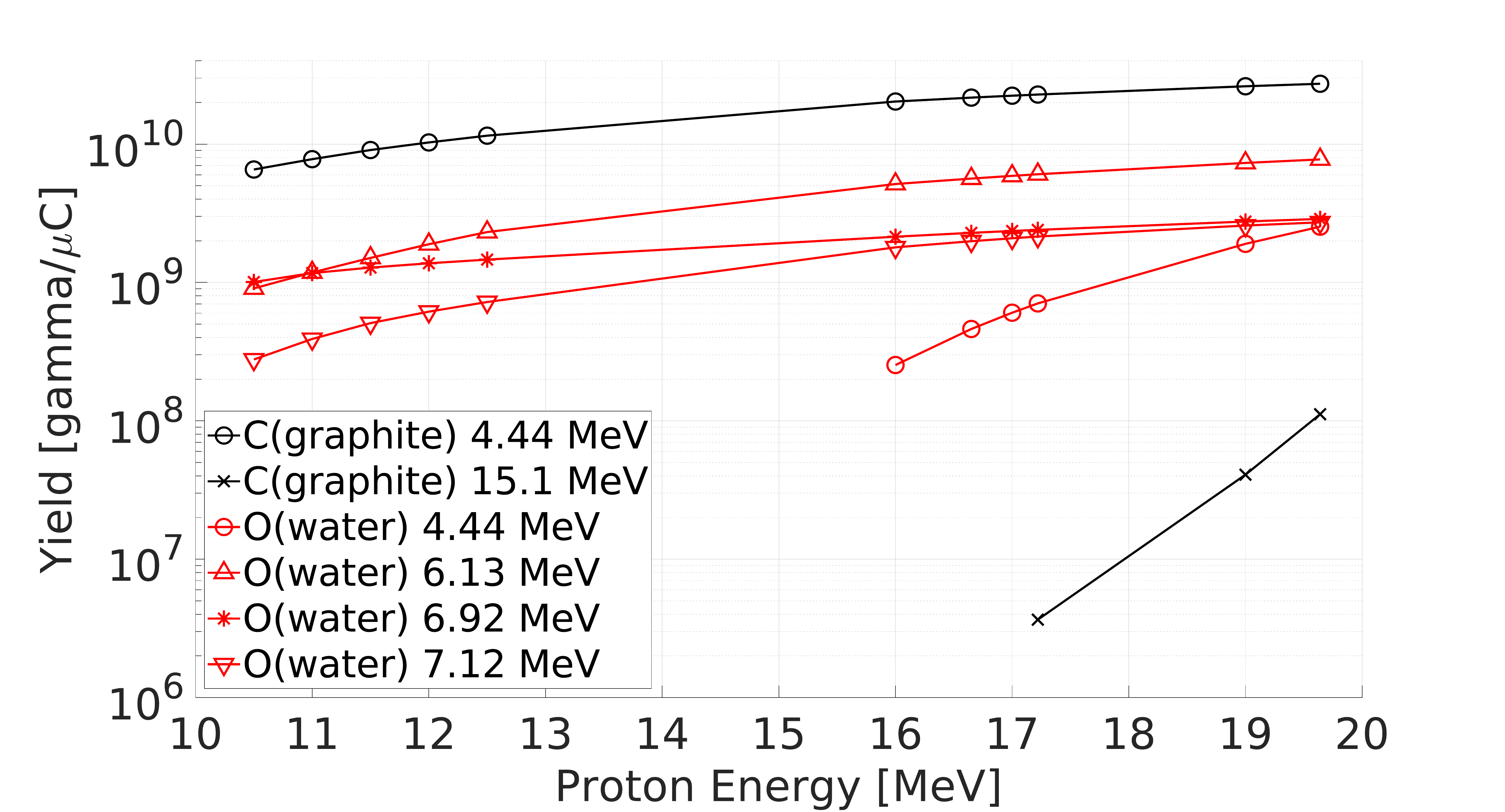}
    \caption{Gamma yield calculations from different $(p,p'\gamma)$ 
    as a function of proton energy. The calculations was performed by numerically integrating $Y = -\frac{\rho N}{A} \int_{E_{th}}^{E_{p}} \dif E [\sigma(E)/{\frac{\dif E}{\dif x}}] $.}
    \label{fig:allyield}
\end{figure}

 The strongest observed photon line is at 4.44~MeV from the $^{12}$C($p,p^{\prime}\gamma$)$^{12}$C reaction. From oxygen (in water), photons are produced by the de-excitation of $^{16}$O$^\ast$ created through ($p,p^{\prime}$) reaction, with energies of 6.13, 6.92, and 7.12~MeV. Using existing cross-section data for the above reactions~\cite{444_angle_cross,all_cross_but151,all_cross_but151_with_angle,444_angle_cross_2,151_cross,151_cross_2} and stopping power data tables~\cite{pstar} for protons, the thick target yields can be calculated by performing a numerical integration using the following equation:
\begin{equation}
Y = \frac{\rho N_A}{A} \int_{E_{th}}^{E_{p}} \dif E \frac{\sigma(E)}{-\frac{\dif E}{\dif x}} 
\end{equation}
 where $E_{th}$ is the threshold energy of corresponding ($p,p^{\prime}\gamma$) reaction, $E_{p}$ is the energy of the protons (12~MeV), $\sigma(E)$ is cross section of the the ($p,p^{\prime}\gamma$) reaction, $\frac{\dif E}{\dif x}$ is the ionization energy loss of the proton, $N_A$ is Avogadro's number, and $\rho$ is the density of the material.  Figure~\ref{fig:allyield} plots the photon yield from different reactions at different proton energies, as determined by the numerical integration.

\subsection{Experimental setup}
\begin{figure}[thb]
    \centering
    \includegraphics[width=\columnwidth]{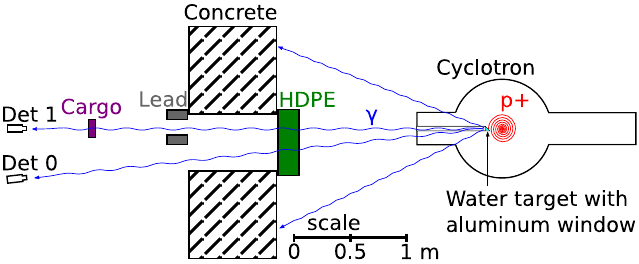}
    \caption{Top view schematic of the MMGR experiment. All the objects on the right of the concrete are in one room while the detectors are in the next room.}\label{fig:bird_eye}
\end{figure}

 A schematic of the experimental setup is shown in Figure~\ref{fig:bird_eye}. Protons are accelerated to approximately 12~MeV, to an orbit  with a radius of 14~cm.  Before striking the internal water target the beam is collimated by a broad graphite collimator. While for most ($p,p^{\prime}\gamma$) reactions the photons are emitted  approximately isotropically, in some reactions there is a factor of two difference in intensity between some emission angles due to a variety of kinematic effects~\cite{444_angle_cross,all_cross_but151_with_angle,613_angle,444_angle_cross_2,151_cross}. Some of the emitted photons then traverse the 5~cm steel wall of the cyclotron and pass through the 10.2~cm $\times$ 83.3~cm opening in the concrete wall.  In order to minimize neutron activation and background radiation outside of the accelerator room, a block of 20.6~cm thick borated high density polyethylene (HDPE) is placed at the opening 251.8~cm from the target.  The photons are further collimated by a 5.1~cm $\times$ 20.3~cm lead collimator. The photon beam then transits through the test object and is measured by detector 1 (Det~1) at a distance of 399~cm away from the proton target. Another detector (Det~0) is positioned 45.7~cm to the side of Det~1, which is equivalent to an angle of 6.54 degrees relative to the photon beam direction. Det~0 thus measures the photons directly from the target, and as such serves as a normalization in the analysis. 
 
 The detectors used to measure the transmitted spectra consist of a 3.81~cm $\times$ 3.81~cm cylindrical LaBr$_3$(Ce) scintillator, allowing high energy resolution of $\sim$3$\%$ (FWHM) and fast processing (0.016~{\textmu}s primary scintillation decay time) at the energies of interest~\cite{canberra,saintgobain}.  The fast scintillation signal from the scintillator combined with the CW proton beam essentially eliminated pileup: the typical count rates in open-air Det 0 were about 10000~s$^{-1}$, using a discriminator threshold that corresponded to $\sim$0.8~MeV. The detector pulses are digitized and processed using CAEN~V1725 waveform digitizer operating in digital pulse processing pulse shape discrimination (DPP-PSD) mode~\cite{caen}. The waveform digitizer is controlled using the ADAQ (AIMS Data AcQuisition) framework, producing data files for subsequent analysis~\cite{adaq}. The  digitizer recorded the arrival time and energy of each pulse to form a transmitted energy spectrum. For both detectors, the energy trigger thresholds are set to approximately 0.8~MeV, filtering the majority of low energy background counts and the 511~keV signals. The integration windows are set at 0.35~{\textmu}s, which allowed for the full capture of the scintillation light pulses from the LaBr$_3$ detector. 
 
\subsection{Experimental Test Objects}
\label{sec: exp cargo and dose}
 A variety of homogeneous and heterogeneous (mixed) materials are used as test objects  in this experiment. The homogeneous materials included aluminum, copper, tin, lead, and depleted uranium, with a range of areal densities. To determine the $Z_{\text{eff}}$ value of the heterogeneous cargo, we first compute the effective $\mu_{\text{eff}}$ at 4.44, 6.13, 6.92, and 7.12~MeV for the mixtures using the following formula:
\begin{equation}
\mu_{\text{eff}}(E) = W_1 \mu_1(E)+ W_2 \mu_2(E)
\label{eq:mu_avg}
\end{equation}
where $W_i$ is the mass fraction of material $i$ in the mixture and $\mu_i(E)$ is the mass attenuation coefficient of material $i$ at energy $E$ as found in XCOM database~\cite{xcom}. The calculated $\mu_{\text{eff}}(E)$ of the mixture are then compared to the list of homogeneous materials in the NIST database at the corresponding photon energy $E$. The $Z$ values of the homogeneous material with the best matching mass attenuation coefficient at the four energies are then averaged and used as the actual  $Z_{\text{eff}}$ for the heterogeneous material.  

Each of the radiography experiments consist of two measurements. The first is a 5 minute run with no material (open air) while the second is a 60 minute run with a test object material. During the analysis, the 60 minute run with the test objects are divided into 10 equal subsets of data.  The reconstruction analysis is performed on individual subsets,  allowing for empirical determination of the fluctuations and accuracy in the reconstructions. Over the course of 19 different experiments, the average current measured on the water target and the graphite collimator are 6.0~{\textmu}A and 0.56~{\textmu}A, respectively. 
 
\section{Analysis}
\label{sec:analysis}
\subsection{Spectral Analysis}
\label{sec:spectral analysis}
\begin{figure}[thb]
    \centering
    \includegraphics[width=\columnwidth]{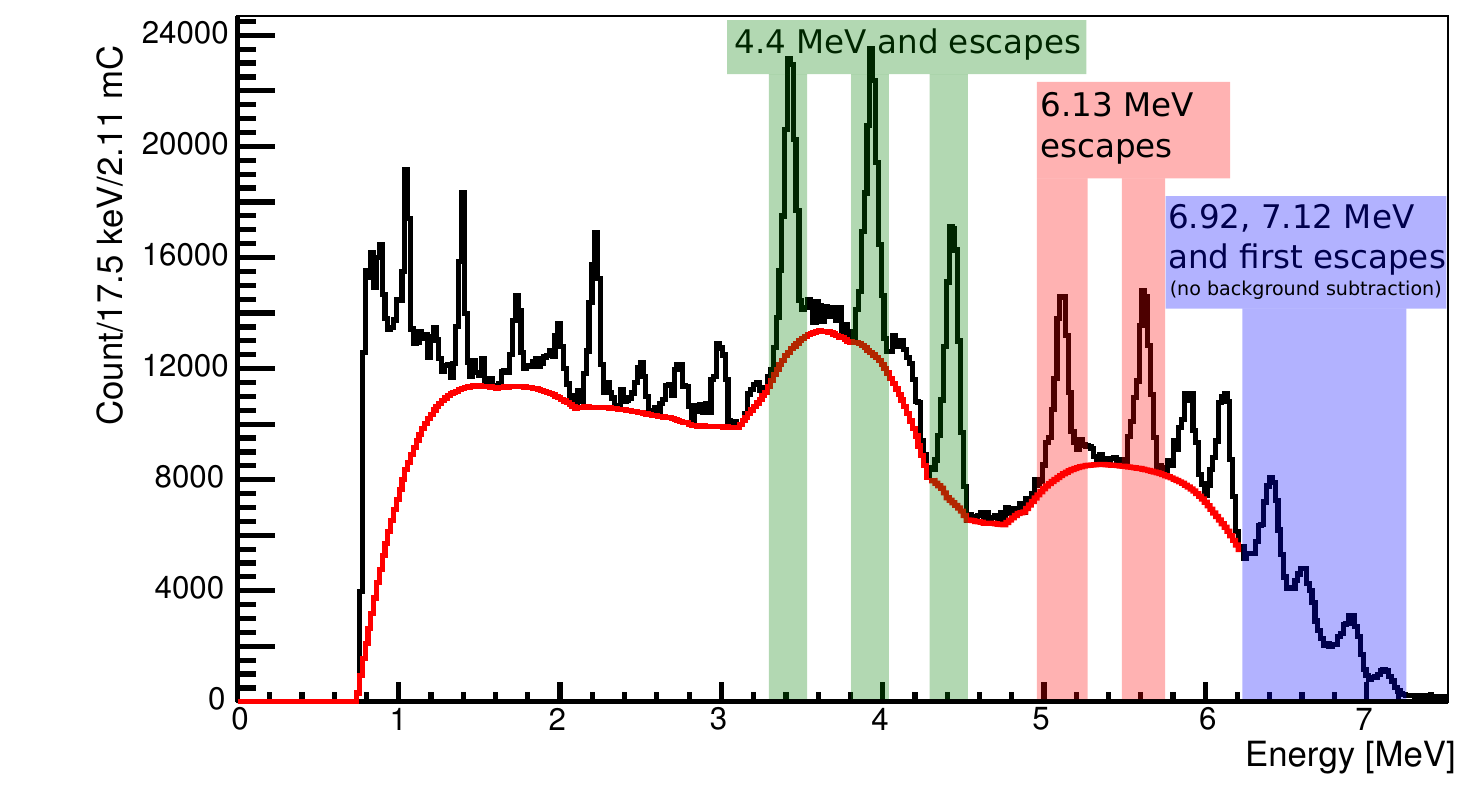}
    \caption{Spectrum of the open beam measured with the on-axis LaBr$_3$(Ce) detector with background fitting plotted in red. The highlighted region are the integration windows for spectral analysis mentioned in Sec.~\ref{sec:spectral analysis}. The bin size for this spectrum is 17.578~keV. The total charge measured on the graphite collimator and the water target are 0.248 and 2.110~mC, respectively. The red line indicates the result of TSpectrum background estimation.}
    \label{fig:openair}
\end{figure}

 The transmission data is analyzed in ROOT, an object oriented data analysis library~\cite{root}. Each spectrum is calibrated using the following procedures: the 2.21, 3.00, 3.42, 3.93, 4.44, 5.11, 5.62, 6.13, 6.41, 6.61, 6.92, and 7.12 MeV peaks are fitted with a Gaussian function using the TFit class. The results of these fits are then used in a second order polynomial calibration fit.  A peak search and background subtraction is then performed on calibrated spectra using the ROOT TSpectrum class.  An example of a resulting spectrum can be seen in Figure~\ref{fig:openair}. 
 The TSpectrum background estimation does not include the 7.12 and 6.92~MeV peaks and their first escapes, as no lines are present above those peaks, making the region encompassing them background-free. For a detailed description of the TSpectrum refer to~\citeauthor{morhavc2000identification}~\cite{morhavc2000identification}.  
 
 After calibration and background subtraction, seven peaks are fitted with a Gaussian at 3.42, 3.93, 4.44, 5.11, 5.62, 6.41, and 7.12~MeV. The numbers of detected counts at 3.42, 3.93, 4.44, 5.11, and 5.62 peaks are determined by tallying counts within two standard deviations from the mean of each fitted peak. The 3.42, 3.93, and 4.44~MeV counts, which constitute the photopeak and escape peaks of the 4.44 MeV photons, are summed as a single effective 4.44~MeV tally. For the 6.13 MeV counts, only the escape peaks at 5.11 and 5.62 MeV are included:  the 6.13 MeV peak consist of both 6.13 MeV photopeak and the 7.12 MeV double escape peak, and there is no simple way of analyzing their contribution individually. An example of the tally regions are highlighted in Figure~\ref{fig:openair}. The region highlighted in purple indicates the summation of total counts in the $6.275 \le E \le 7.242$~MeV region. These counts originated from the 6.92 and 7.12~MeV photons alone. 

\subsection{Reconstruction Algorithm}
\begin{figure}[thb]
    \centering
    \includegraphics[width=\columnwidth]{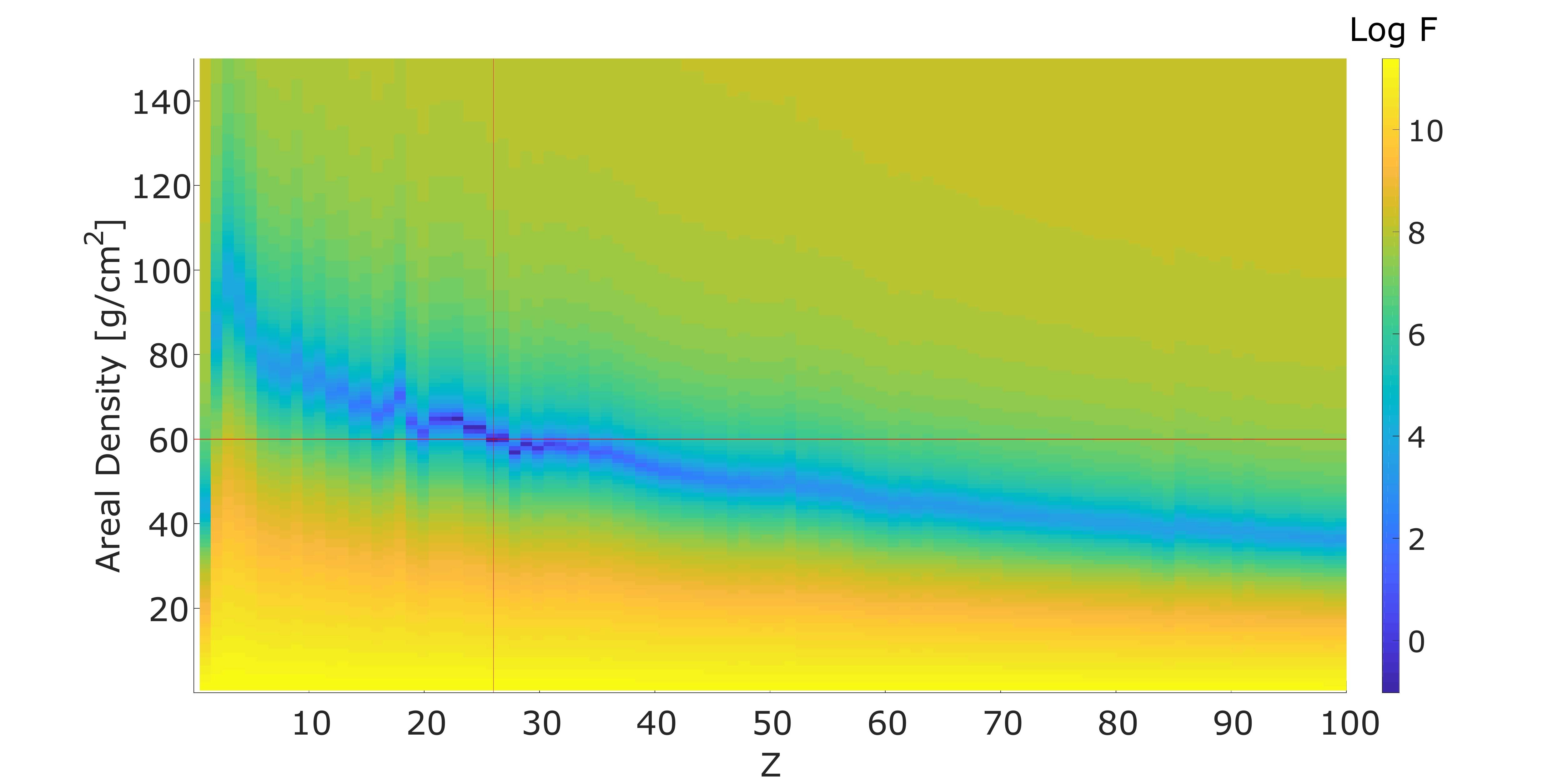}
    \caption{Heatmap of the reconstruction metric $F$ as a function of $Z$ and $x$. The test object material in this experiment is 60.3 g/cm$^2$ of copper ($Z$ = 29). The red line indicate the predicted $Z_{\text{eff}}$ and $x$ estimation of 26 and 60 g/cm$^2$, respectively.}
    \label{fig:heamap}
\end{figure}

 After obtaining the transmitted counts from the spectra in the three regions of interest, the counts measured by Det~1 (on-axis) is first normalized by counts in Det~0 (off-axis) as follows to minimize the various time-dependent systematic effects: $C(E) = \frac{C_{1}(E)}{C_{0}(E)}$,  where $C_{0}(E)$ and $C_{1}(E)$ are the recorded counts from Det~0 and Det~1 at energy $E$. These systematic effects could originate from beam focusing, vacuum quality, and the temperature of internal parts of the cyclotron ultimately leading to variations in beam position and energy, which in its turn leads to changes in photon yields over time. The normalized counts are then used to determine the transmission ratio:
\begin{equation}
R_\text{exp}(E) = \frac{C_{\text{cargo}}(E)}{C_{\text{air}}(E)}
\label{eq:expratio}
\end{equation}
 where $C_{\text{cargo}}(E)$ and $C_{\text{air}}(E)$ are the normalized counts at energy $E$ for the test object and open beam.  This ratio also achieves the cancellation of geometry-dependent systematic effects.  As a key step of the reconstruction this ratio is then compared to calculated ratios predicted from a variety of possible values of areal density and $Z$. For $Z$ of 1 to 100 and areal density $x$ of 1 to 150 g/cm$^2$ a table of ratios $R_\text{calc}(E,Z,x)$ is calculated at each combination of $Z$ and $x$ using Eq.~\ref{eq:total_attenuation} and table of $\mu$ from~\citeauthor{xcom}\cite{xcom}.  With the computed table of attenuation and measured transmission rations at different energies, a figure of merit $F$ is defined: 
 \begin{equation}
\begin{aligned}[b]
F = \sum_{i = 1}^{2}\left\{ \frac{\frac{R_\text{exp}(E_i)}{R_\text{exp}(7.017)}-\frac{R_\text{calc}(E_i,Z,x)}{R_\text{calc}(7.017,Z,x)}}{\sigma\left[\frac{R_\text{exp}(E_i)}{R_\text{exp}(7.017)}-\frac{R_\text{calc}(E_i,Z,x)}{R_\text{calc}(7.017,Z,x)} \right]} \right\}^2 \\
-\left\{\frac{R_\text{exp}(7.017)-R_\text{calc}(7.017,Z,x)}{\sigma\left[ R_\text{exp}(7.017)-R_\text{calc}(7.017,Z,x) \right]}\right\}^2
\label{eq:full eq}
\end{aligned}
\end{equation}
 where $E_1$ is 4.44~MeV and $E_2$ is 6.13~MeV, and $\sigma$ represents the statistical uncertainty of the numerator, as determined by a propagation of errors returned by the TSpectrum class.  Here $R(7.017)$ and similar refer to the count ratios for the $^{16}$O 7.016~MeV line. The summation term is primarily sensitive to $Z_{\text{eff}}$. The second term is additionally sensitive to the areal density.  The process of the reconstruction then searches for a global minimum in $F$ space and identifies the corresponding values of $Z$ and $x$ as the reconstructed values.  Figure \ref{fig:heamap} shows a heat map of calculated $F$ for the case of a copper cargo with areal density of 60.3 g/cm$^2$, with reconstructed values of $Z_{\text{eff}}=26$ and $x=60$~g/cm$^2$ very close to the actual numbers.

\section{Results}
\label{sec:res}
\begin{figure}[!ht]
    \centering
    \includegraphics[width=\columnwidth]{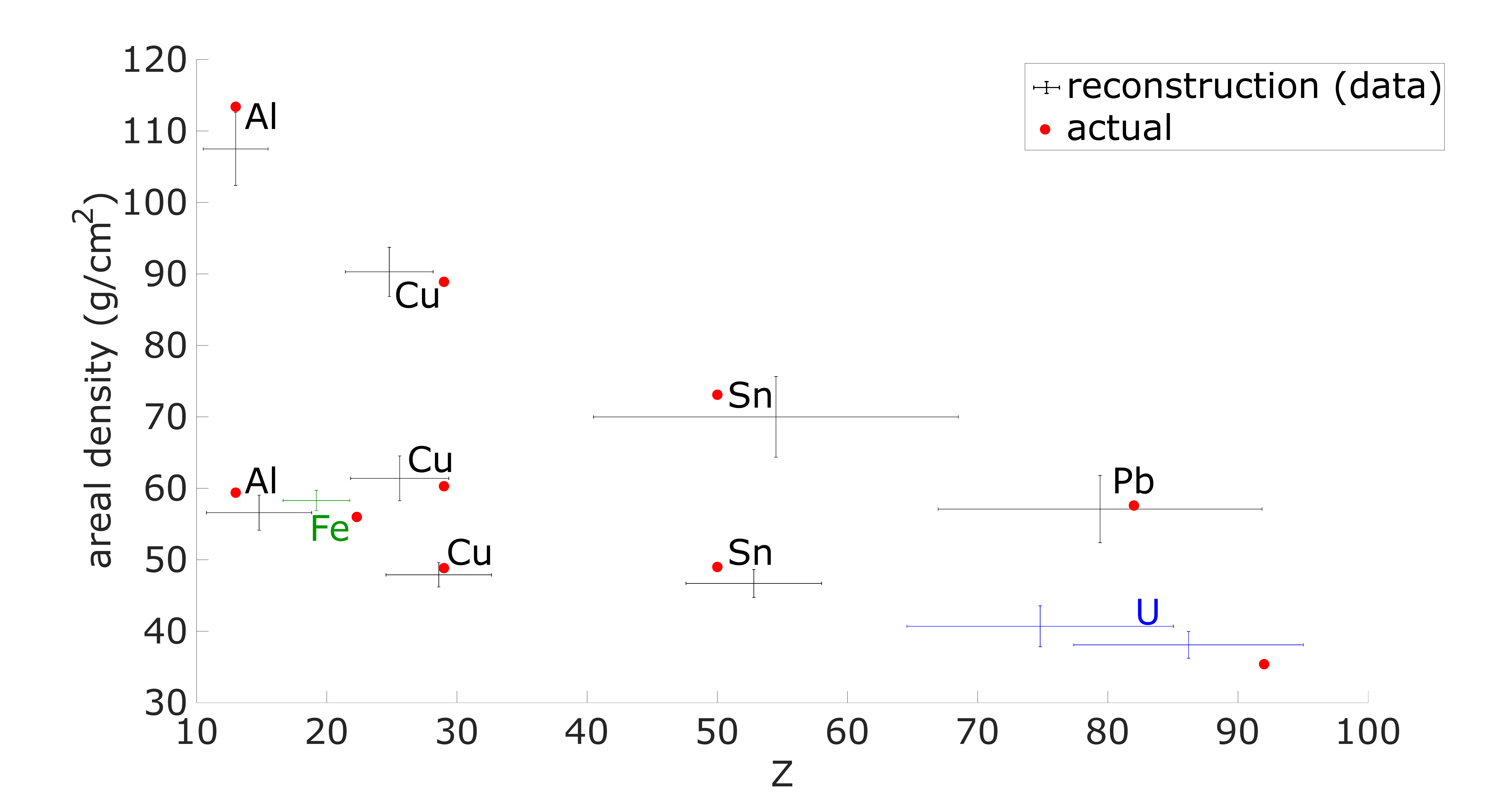}
    \includegraphics[width=\columnwidth]{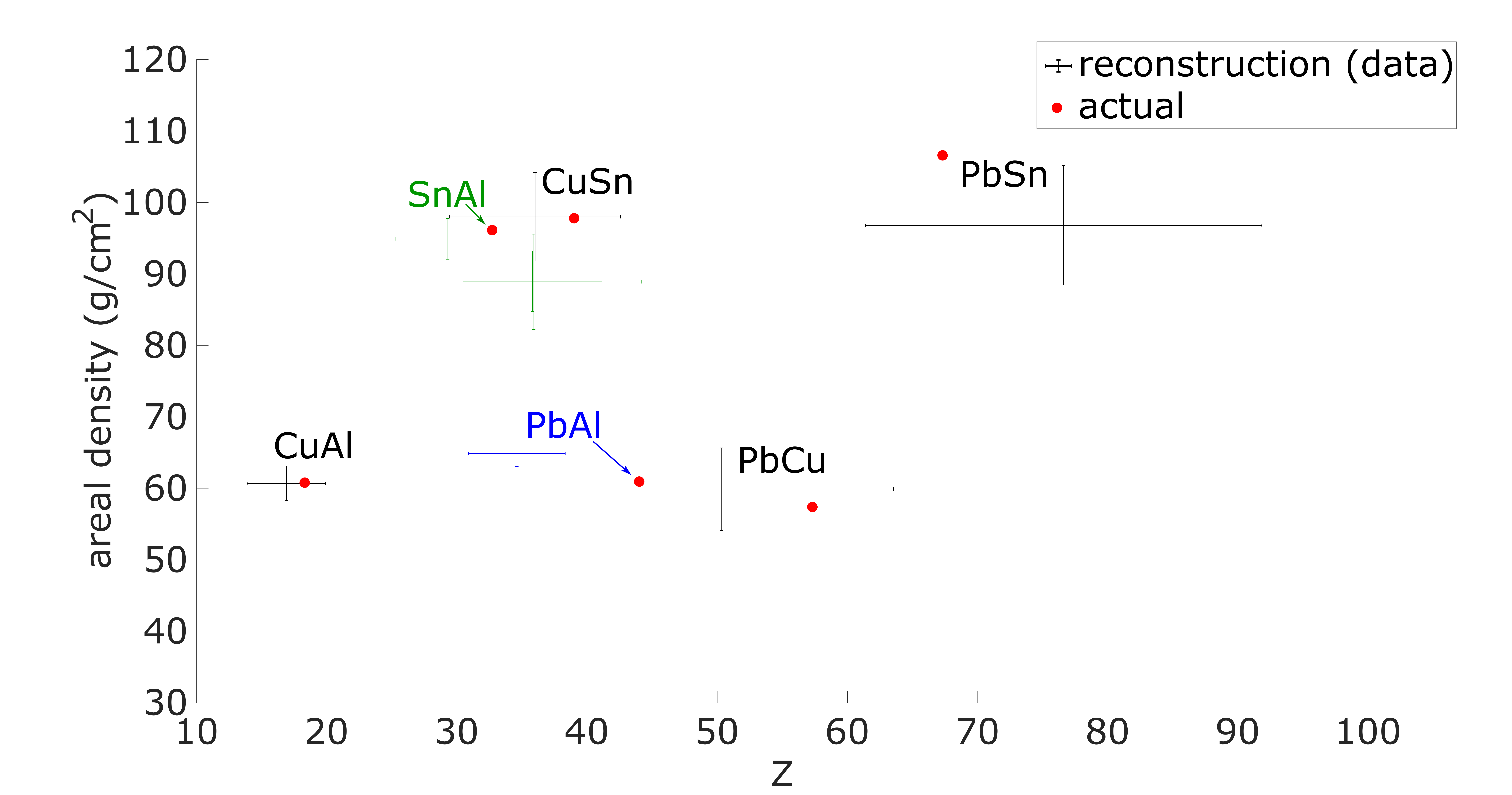}

    \caption{Reconstruction results of all homogeneous material (top) and heterogeneous material (bottom) MMGR experiments. The red dots indicating the actual $Z$ and $x$ value of the materials. The experimental points show the mean and standard deviation over the $Z_{\text{eff}}$ and $x$ in the data set for a particular test object. Two separate experiments were performed on the same depleted uranium test object, and three separate experiments were performed on the same Sn+Al heterogeneous test object.}
    \label{fig:results}
\end{figure}
 As mentioned in Sec.~\ref{sec: exp cargo and dose}, each of the 60 minute experimental transmission datum are equally divided into ten smaller data subsets. The radiography analysis described in Sec.~\ref{sec:spectral analysis} is then performed on each of the data sets. The mean and standard deviation of the reconstructed $Z$ and $x$ are computed across the the data set, and these are compared to the actual known values of the test objects. These reconstruction results are tabulated in Table~\ref{tab:results} and are plotted in Figure~\ref{fig:results} for homogeneous and heterogeneous cargoes. In most cases the reconstructed values are within the statistical uncertainty of the actual values.  
The large size of the error bars for the higher $Z$ and higher $x$ materials reflects the limited statistics of the transmitted counts, due to increased attenuation at those values.  It is now possible to compare the reconstructed and actual values of $Z_{\text{eff}}$ and $x$ in a $\chi^2$ test, as a way to determine whether the deviations are due to statistics or some residual, unaccounted-for systematic effects. The reduced $\chi^2$ of $Z_{\text{eff}}$ and $x$ between the average reconstructed values and the actual values for all 19 experiments are 0.90 and 1.28, respectively. These correspond to p-values of 0.58 and 0.19, respectively, indicating the deviations are dominated by statistics. However, as the standard deviation of the reconstructed values were driven by experimental data calculated from the reconstruction values, systematic error such as cyclotron instability may still contribute a large part. 
As seen in Figure \ref{fig:results}, there is significant separation between low-$Z$ material such as aluminum, medium-$Z$ material such as tin, and high-$Z$ material such as lead. Furthermore, there is a distinct separation between materials with different $x$. The difference of reconstructed $Z$ between lead and uranium are small, however, partially due to the insufficient photon transmission statistics in the tests with high-$Z$ materials.

\section{Conclusions and Future Work}
 
 In the experiments presented in this study we demonstrated the feasibility of using  monoenergetic photons generated from nuclear reactions to  reconstruct the areal density and $Z_{\text{eff}}$ of a variety of homogeneous and heterogeneous test objects with a broad range of $Z$. 
 The precision of the measurements was limited by the statistical precision of the collected data, in part explaining the lack of separation between the uranium and lead reconstructions in Figure~\ref{fig:results}.  Future work should focus on using  accelerators with higher energies, as this will increase the yield of the photons, as shown in calculations described in Figure~\ref{fig:allyield}.  For example, increasing the proton beam energy from 12~MeV to 19~MeV would more than double the photon yields, while being only slightly above the neutron production thresholds.  While the experiment used beam currents of 6~{\textmu}A, higher beam currents are necessary to achieve a discrimination between lead and uranium:   based on the results in this study we estimate that a 300~{\textmu}A, 12~MeV proton beam is required for a 5$\sigma$ discrimination between Pb and U.  Additionally, the internal target of the accelerator presents a significant limitation. Future research should focus on the development of cyclotrons with external beams while achieving better beam stability.  Facilities with these characteristics already exist, such as the IBA C18/9 cyclotron at A. Alikhanyan National Laboratory used for radioisotope production~\cite{danagulyan2016some, avagyan2018study}. Furthermore, research to develop more compact cyclotrons with these specifications is already underway by~\citeauthor{antaya2018isochronous}\cite{antaya2018isochronous}, and will allow for less shielding, eliminating the need for the concrete shielding used in this experiment.  It should be noted that other methods for producing monoenergetic gammas exist as well, e.g. through Thompson scattering laser methods\cite{geddes2015compact}.  The main advantage of the cyclotron-based methods is in the compactness of the platform, while the advantage of the laser methods is in the tunability of the source's energy.  Finally, advances in high temperature superconductors may in the future allow for cyclotrons with stronger magnetic fields, leading to smaller sizes, lower cost, and thus increased applicability.
 

\begin{acknowledgments}
This work was supported in part by the U.S. Department of Homeland Security Domestic Nuclear Detection Office under a competitively awarded collaborative research ARI-LA Award, ECCS-1348328. B.S.H. gratefully acknowledges the support of the Stanton Foundation Nuclear Security Fellowship.  The research also benefited   from generous funding by Andiscern Corporation, which additionally provided the accelerator used in the study. The authors are grateful to Richard C. Lanza, who developed some of the initial ideas behind this work, for his support, encouragement, and valuable advice. The authors wish to thank Steven J. Jepeal and Prof. Zachary Hartwig for the collaborative effort involving the ION-12$^{\text{SC}}$ that made this work possible.
\end{acknowledgments}

The data that support the findings of this study are available from the corresponding author upon reasonable request.

\begin{table*}[!htbp]
\begin{tabular}{l r r r | r r r}
\hline\hline \\
Material & Actual~~~& Average Reconstructed~~~~& $\sigma _{Z_{\text{eff}}}$ & Actual~~~& Average Reconstructed  & $\sigma _x$ \\
         &     $Z_{\text{eff}}$          & $Z_{\text{eff}}$ &  & $x$ (g/cm$^2$) &  $x$ (g/cm$^2$)     &    (g/cm$^2$)       \\
\hline \\

Aluminum 1 (Al) & 13          & 14.8   & 4.0   & 59.4  & 56.6  & 2.5  \\
Aluminum 2 (Al) & 13          & 13.0   & 2.5   & 113.4 & 107.5 & 5.1  \\
Iron (Fe)       & $\sim$22    & 19.2   & 2.6   & 56.0  & 58.3  & 1.4  \\
Copper 1 (Cu)   & 29          & 28.6   &  4.1  & 48.8  & 47.9  & 1.7  \\
Copper 2 (Cu)   & 29          & 25.6   &  3.8  & 60.3  & 61.4  & 3.1  \\
Copper 3 (Cu)   & 29          & 24.8   &  3.4  & 88.9  & 90.3  & 3.4  \\
Tin 1 (Sn)      & 50          & 51.6   &  5.0 & 49.0  & 47.1  & 1.8  \\
Tin 2 (Sn)      & 50          & 54.5   &  14.0 & 73.1  & 70.0  & 5.6  \\
Lead (Pb)       & 82          & 79.4   &  12.4 & 57.6  & 57.1  & 4.7  \\ 
Uranium (U)     & 92          & 86.2   &  8.8  & 35.4  & 38.1  & 1.9  \\ 
Uranium (U)     & 92          & 74.8   &  10.2 & 35.4  & 40.7  & 2.9  \\ 
Pb $+$ Al       & $\sim$44    & 34.6   &  3.7  & 61.0  & 64.9  & 1.9  \\
Cu $+$ Al       & $\sim$18    & 16.9   &  3.0  & 60.8  & 60.7  & 2.4  \\
Pb $+$ Cu       & $\sim$57    & 50.3   &  13.2 & 57.4  & 59.9  & 5.8  \\
Pb $+$ Sn       & $\sim$67    & 76.6   &  15.2 & 106.6 & 96.8  & 8.4  \\
Cu $+$ Sn       & $\sim$39    & 36.0   &  6.6  & 97.8  & 98.0  & 6.2  \\
Sn $+$ Al       & $\sim$33    & 35.9   &  8.3  & 96.1  & 88.9  & 6.7  \\
Sn $+$ Al       & $\sim$33    & 29.3   &  4.0  & 96.1  & 94.9  & 2.8  \\
Sn $+$ Al       & $\sim$33    & 35.8   &  5.3  & 96.1  & 89.0  & 4.2  \\
\hline\hline
\end{tabular}
\caption{Average reconstructed $Z_{\text{eff}}$ and areal density $x$ for all MMGR experiments with actual values and reconstruction uncertainties listed. The $\sim$ symbol indicates the actual $Z$ value of the heterogeneous cargo calculated with the method as described in Sec.~\ref{sec: exp cargo and dose}.}
\label{tab:results}
\end{table*}



\bibliography{references}

\end{document}